\documentclass{czjphys}         
\usepackage{epsf,cite}

\begin{document}
\title{Resonance expansions in quantum mechanics}  
%

\authori{Rafael de la Madrid}      
\addressi{Departamento de F{\'\i}sica 
Te\'orica, Facultad de Ciencias, Universidad del Pa{\'\i}s Vasco, E-48080 
Bilbao, Spain \\  E-mail:  {\texttt{wtbdemor@lg.ehu.es}} }

\authorii{Gast\'on Garc{\'\i}a-Calder\'on}     
\addressii{Instituto de F{\'\i}sica, Universidad Nacional Aut\'onoma
de M\'exico, 01000 M\'exico DF, M\'exico}

\authoriii{Juan Gonzalo Muga}    
\addressiii{Departamento de 
Qu{\'\i}mica-F{\'\i}sica, Facultad de Ciencias, Universidad del Pa{\'\i}s 
Vasco, E-48080 Bilbao, Spain}

\authoriv{}     \addressiv{}

%
\headauthor{R.~de la Madrid, G.~Garc{\'\i}a-Calder\'on, and J.G.~Muga} 
\headtitle{Resonance expansions in quantum mechanics}  
\lastevenhead{Resonance expansions} 
\pacs{3.65-w}     

\refnum{6007 C} 
\daterec{29 July 2005}    
\issuenumber{9}  \year{2005}
\setcounter{page}{1}

\maketitle

\begin{abstract}
\noindent The goal of this contribution is to discuss various resonance 
expansions that have been proposed in the literature.
\end{abstract}

\section{Introduction}
\setcounter{equation}{0}

The Gamow (or resonance) states are the wave functions of 
resonances. These states are eigensolutions of the Schr\"odinger equation 
subject to a ``purely outgoing boundary condition.'' The Gamow states
were introduced by Gamow~\cite{GAMOW} in 1928 to describe $\alpha$ decay (see 
also~\cite{CONDON}). Some years later, in 1939, Siegert made use of the 
Gamow states to obtain a resonance expansion of the 
scattering function for potentials of finite range~\cite{SIEGERT} which was 
further developed by Humblet 
and Rosenfeld \cite{HUMBLET}. In 1955, Peierls pointed out the relationship 
between the residues of the propagator at the resonance poles and the Gamow 
states~\cite{PEIERLS55}. The properties of the Gamow states and 
its applications have been considered by many authors, see for 
example~\cite{PEIERLS59,ZELDOVICH,BERGGREN,gcp76,ROMO,BG,LIND,VERTSE,BOLLINI,
TOLSTIKHIN,FERREIRA,MONDRAGON,CIVITARESE,MICHEL,SANTRA}. A pedestrian 
introduction to these states can be found in~\cite{AJP02}.

The Gamow states can be used to obtain resonance expansions of wave functions 
and propagators. The purpose of this paper is to discuss the main features of 
several of such expansions and to point out some of their differences. For 
the sake of simplicity, we shall restrict our discussion to potentials of 
finite range. The potentials will be three dimensional and spherically 
symmetric except in Section~\ref{sec:stdesc}, where they will be assumed to be 
one dimensional.

\section{Berggren's and Berggren-like resonance expansions}

\begin{figure}[ht]
\begin{center}
\epsfxsize=10cm
\epsfbox{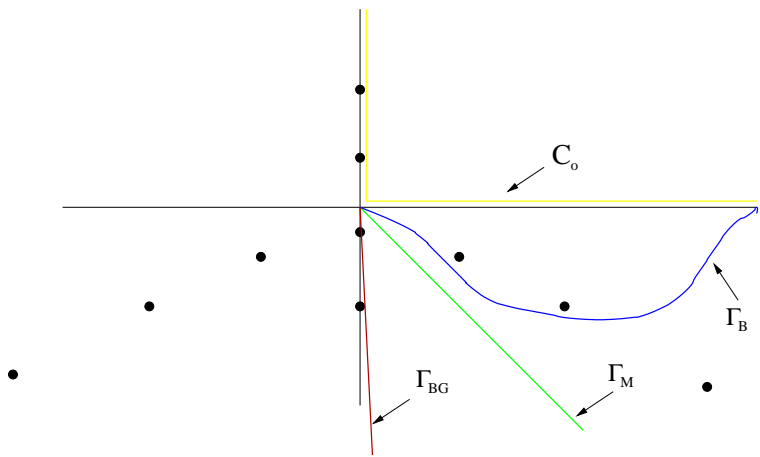}
\end{center}
\caption{\label{contours}Different contours yield different expansions. The
bullets represent the bound, resonance, anti-resonance and virtual poles.}
\end{figure}

In order to derive Berggren's expansion~\cite{BERGGREN}, we start out with a 
completeness relation in terms of the bound and scattering states,
\begin{equation}
     1=\sum_{n=1}^{N_{\rm b}} |K_n\rangle \langle K_n| +
       \int _{0}^{\infty}\D k \,  |k^+\rangle \langle ^+k| \, ,
         \label{resonidse}
\end{equation}
where $1$ is the identity operator, $|K_n\rangle$ are the bound states and 
$|k^+\rangle$ are the scattering states. We are assuming that the potential 
holds ${N_{\rm b}}$ bound states. By deforming the integral of 
Eq.~(\ref{resonidse}) into the contour ${\Gamma}_{\rm B}$ of 
Fig.~\ref{contours}, and by using Cauchy's theorem, we extract the 
contributions from the resonance states that are hidden in the continuum 
integral and write them in the same form as the bound state contributions:
\begin{equation}
     1=\sum_{n=1}^{N_{\rm b}} |K_n\rangle \langle K_n| +
        \sum_{n=1}^{N_{\rm B}}  |k_n^+\rangle \langle ^+k_n| 
       + \int _{\Gamma _{\rm B}}\D k \,  |k^+\rangle \langle ^+k| \, ,  
         \label{resonidBerggren}
\end{equation}
where $|k_n^+\rangle$ is the Gamow state corresponding to the
$n$th resonance, and $N_{\rm B}$ is the number of resonances that lie
in between the real axis and the contour $\Gamma _{\rm B}$. The integral
in Eq.~(\ref{resonidBerggren}) is called the background.

Berggren's expansion includes only those few resonances that are supposed to 
be the most important in the energy range of interest. The background 
integral is usually assumed to give a negligible contribution in such energy 
range. However, the background need not always be negligible. For example, 
virtual poles influence low energy scattering, and the background integral 
in Eq.~(\ref{resonidBerggren}) cannot be neglected at low energies. (Of 
course, one can easily deform the contour $\Gamma _{\rm B}$ to include 
explicitly the effect of the virtual poles.) As well, even when the 
background integral is negligible, it is never
zero. This is due to the fact that the bound and resonance states do not 
form a complete basis, and one has to include an additional sets of kets to
obtain a complete basis. When the background integral is not negligible,
which is the usual case encountered in applications, one needs to estimate 
it. For Berggren's expansion, a discretization of the background integral 
yields such estimation~\cite{FERREIRA}.

One can construct Berggren expansions of the Hamiltonian $H$, the resolvent 
$1/(z-H)$ and the evolution operator $\E ^{-\I Ht}$ by simply 
letting those operators act on Eq.~(\ref{resonidBerggren}).

Berggren's expansion~(\ref{resonidBerggren}), sometimes with slight
modifications, is the one most oftenly used in nuclear 
physics~\cite{LIND,VERTSE,FERREIRA,MICHEL}.

Needless to say, the completeness relation~(\ref{resonidBerggren}) is a formal
expression that must be understood within the rigged Hilbert space as part
of a ``sandwich'' with well-behaved wave functions $f$ and $g$:
\begin{equation}
     (f,g)=\sum_{n=1}^{N_{\rm b}} \langle f|K_n\rangle \langle K_n|g\rangle +
          \sum_{n=1}^{N_{B}} \langle f|k_n^+\rangle \langle ^+k_n|g\rangle +
 \int _{\Gamma _{B}}\D k \,  \langle f|k^+\rangle \langle ^+k|g\rangle 
          \, .
         \label{resonidsefg}
\end{equation}
Expression~(\ref{resonidsefg}) is
valid only when the wave functions $f(r)$ and $g(r)$ fall off at infinity 
faster than any exponential. 

As we mentioned above, Berggren's expansion includes only the resonances that
carry the most influence in the energy range under consideration. One can 
incorporate the contribution of other resonances by enclosing other poles 
of the fourth quadrant of the $k$-plane. For example, by
using the contour $\Gamma _{\rm M}$ of Fig.~\ref{contours}, one 
obtains~\cite{MONDRAGON}
\begin{equation}
     1=\sum_{n=1}^{N_{\rm b}} |K_n\rangle \langle K_n| +
        \sum_{n=1}^{N_{\rm M}}  |k_n^+\rangle \langle ^+k_n| 
       + \int _{\Gamma _{\rm M}}\D k \,  |k^+\rangle \langle ^+k| \, ,  
         \label{resonidMondra}
\end{equation}
where $N_{\rm M}$ is the number of (proper) resonances in between the 
contour $\Gamma _{M}$ and the real axis. By substituting 
$|k^+\rangle = S(k)|k^-\rangle$ into Eq.~(\ref{resonidse}), where $S(k)$ is 
the $S$ matrix and $|k^-\rangle$ is the ``out'' Lippmann-Schwinger ket, and 
by using the contour $\Gamma _{\rm BG}$ of Fig.~\ref{contours}, one 
obtains another expansion~\cite{BG}:
\begin{equation}
     1=\sum_{n=1}^{N_{\rm b}} |K_n\rangle \langle K_n| +
        \sum_{n=1}^{N_{\rm BG}}  |k_n^-\rangle \langle ^+k_n| 
       + \int _{\Gamma _{\rm BG}}\D k \,  |k^-\rangle S(k) \langle ^+k| \, ,  
         \label{resonidBG}
\end{equation}
where $N_{\rm BG}$ is the number of resonances in the fourth quadrant of
the complex plane. One can also include virtual states in an obvious way.

Likewise the completeness relation~(\ref{resonidBerggren}), 
the completeness relations~(\ref{resonidMondra}) and (\ref{resonidBG}) are 
to be understood as part of a ``sandwich'' with well-behaved functions
$f$ and $g$. However, unlike 
the completeness relation~(\ref{resonidBerggren}), the completeness
relations~(\ref{resonidMondra}) 
and (\ref{resonidBG}) do not make sense as they stand. The reason is that, 
in order to properly derive those expansions, it is necessary that the analytic
continuation of the integrands $\langle g|k^+\rangle \langle ^+k|f\rangle$ and 
$\langle g|k^-\rangle S(k) \langle ^+k|f\rangle$
tends to zero as $k$ tends to infinity in the fourth quadrant of the
complex plane. However, those integrands diverge exponentially in that
limit. For example, if $f(r)$ is an infinitely differentiable
function with compact support, then its wave number representation
$f(k)=\langle ^+k|f\rangle$ diverges exponentially on the infinite 
arc. Therefore, one has to use either a regulator or a time-dependent 
approach:
\begin{equation}
     \E ^{-\I Ht}=
      \sum_{n=1}^{N_{\rm b}} \E ^{-\I K_n^2t} |K_n\rangle \langle K_n| +
        \sum_{n=1}^{N_{\rm BG}} \E ^{-\I k_n^2t} |k_n^+\rangle \langle ^+k_n| 
        + \int _{\Gamma _{\rm M}}\D k \, \E ^{-\I k^2t}
              |k^+\rangle \langle ^+k| \, , \quad t>0 \, .  
         \label{resonidCaval}
\end{equation}
The expansions~(\ref{resonidMondra}) and (\ref{resonidBG}) are now
understood as the (singular!) limit when $t\to 0$ of the 
expansion~(\ref{resonidCaval}). Note that the expansion~(\ref{resonidCaval})
is valid for $t>0$ only, yet another reminder that resonances 
ought to be understood in a time-asymmetric, time-dependent manner. We shall
further discuss time-dependent expansions in the next section.

A side effect of the divergence of the integrand
$\langle g|k^-\rangle S(k) \langle ^+k|f\rangle$ at the infinite arc is that
the so-called ``Hardy axiom''~\cite{BOHM,BG} is flawed. The reason
is that the ``Hardy axiom'' assumes that $\langle f|k^-\rangle$ and 
$\langle ^+k|g\rangle$ are Hardy functions, and therefore 
$\langle f|k^-\rangle$ and $\langle ^+k|g\rangle$ should tend to zero in the 
infinite arc of the fourth quadrant of the complex $k$-plane. On the contrary,
the quantum mechanical wave functions $\langle f|k^-\rangle$ and 
$\langle ^+k|g\rangle$ diverge (exponentially!) in that limit, and therefore
they cannot be Hardy functions.

\section{Time-dependent expansions}

\subsection{Expansion involving proper resonance poles}

As is well known, the solution $\psi(t)$ to the time-dependent 
Schr\"odinger equation may be written in terms of the retarded time evolution 
operator ${\rm exp}(-\I Ht)$, where $ t \geq 0$, and of the known arbitrary 
initial state $\psi(0)$ as
\begin{equation}
       \psi(t) = \E^{-\I Ht} \psi(0)   \, .
        \label{3}
\end{equation}
In coordinate representation, the retarded Green function 
$g(r,r';t)=\langle r|\E ^{-\I Ht}|r'\rangle$ may be written, using
the Laplace transform, as~\cite{gcp76}
\begin{equation}
    g(r,r';t)={\I \over 2 \pi} \int_{C_{\circ}}
    \D k \,  G^+(r,r';k) \, \E ^{-\I k^2t} \,2k  \, , 
\label{4}
\end{equation}
where $G^+(r,r';k)$ denotes the outgoing, time-independent Green function. The
contour $C_{\circ}$ runs parallel to the positive imaginary $k$-axis and bends
close to the origin to run parallel to the positive real $k$-axis, staying 
always in the first quadrant of the complex $k$-plane, see 
Fig.~\ref{contours}. It 
is possible to expand the time-dependent Green function in terms of the
resonance states plus a background integral by deforming appropriately the 
contour $C_{\circ}$ in the $k$-plane~\cite{gcp76}. Since the variation 
of $G^+(r,r';k)$ with $k$ complex is at most exponential~\cite{newton}, the 
behavior of the integrand with $k$ in Eq.~(\ref{4}) is dominated by 
exp$(-\I k^2t)$. A convenient choice is to deform $C_{\circ}$ to a contour 
involving two semi-circles $C_s$ along the second and fourth quadrants of the
$k$-plane plus a straight line $C_L$ that passes through the origin at 
$45^{\circ}$ off the real $k$-axis. In doing so, one passes over some poles 
of the outgoing Green function. In general, these include bound states and 
the subset of complex poles associated with the so-called proper resonance 
states, i.e., those poles for which ${\rm Re}\, k_p > {\rm Im} \,k_p$. By 
extending the integration 
contour up to infinity, one obtains that the semi-circles $C_s$ yield
a vanishing contribution, so one is left with a infinite sum of terms
plus an integral contribution~\cite{gcp76}:
\begin{eqnarray}
     g(r,r';t) &=& 
       \sum_{p=1}^{\infty} u_p(r)u_p(r')\E ^{-\I k_p^2t}
               \nonumber\\
   &&+{\I \over 2\pi} \int_{C_L} \D k \, G^+(r,r';k) \, \E^{-\I k^2t}\,2k\, ,
        \label{3.6}
\end{eqnarray}
where, without loss of generality and for the sake of simplicity of the 
expressions, we have omitted the bound states. In the last equation, the 
sum runs through the proper resonance poles --hence the subscript $p$-- of 
the outgoing Green function. In deriving expression~(\ref{3.6}), 
one uses that the residue of $G^+(r,r';k)$ at the complex pole $k_p$ is 
$u_p(r)u_p(r´)/2k_p$, and that the resonance states are normalized according 
to the following condition~\cite{gcp76}:
\begin{equation}
        \int_0^R \D r \,  u_p^2(r) + \I \frac {u_p^2(R)}{2k_p} =1 \, ,
           \label{2d}
\end{equation}
where $R$ denotes the radial coordinate from which the potential 
vanishes. Note that the integral term in Eq.~(\ref{3.6}) may be written as an 
integral along the contour $\Gamma _{\rm M}$.

\subsection{Expansion involving the full set of poles and resonance states}

One may obtain an expansion involving the full set of bound, virtual,  
resonance and anti-resonance states (the latter associated with the 
third-quadrant poles of the $S$ matrix) by noting that the 
contour in Eq.~(\ref{4}) may be deformed into a 
semi-circle $C_s'$ along the third quadrant of the $k$-plane plus an integral 
term along the real $k$-axis. By extending the integration contour up to 
infinity, we again obtain that the contribution of the semi-circle vanishes
and we are left with the expression
\begin{equation}
     g(r,r';t)={\I \over 2 \pi} \int_{-\infty}^{+\infty} \D k \, 
         G^+(r,r';k) \E^{-\I k^2t} \,2k \, .
      \label{4.1}
\end{equation}
It turns out that under the condition  $r,\, r' < R$,
one may use the Cauchy expansion of the outgoing Green function
\begin{equation}
       G^+(r,r';k) = \frac{1}{2k}\sum_{n=-\infty}^{\infty} 
        \frac {u_n(r)u_n(r')}{(k-k_n)} \, , \hskip.9truecm  r,\,r' < \, R \, ,
        \label{15}
\end{equation}
where the anti-resonance poles correspond to negative integer values of 
$n$. The above expression still holds when either $r$ or $r'$, but not 
both, equals $R$. To our knowledge, the above result was first reported, for 
a solvable model, by More~\cite{more}. It has been proved in 
the s-wave case for potentials of finite range by Garc\'{\i}a-Calder\'on and 
Berrondo~\cite{gcb79}, and it has been used by several authors, see 
for example~\cite{gareev,gc82,gcr86}. The expansion given by Eq.~(\ref{15}) 
implies that the Gamow states satisfy the relationships
\begin{equation}
    \sum_{n=-\infty}^{\infty}\frac{u_n(r)u_n(r')}{k_n}=0 \, ,
        \hskip.9truecm  r,r'<R \, , 
    \label{12}
\end{equation}
and
\begin{equation}
        \frac{1}{2}\sum_{n=-\infty}^{\infty} u_n(r)u_n(r')=\delta (r-r') \, , 
          \hskip.9truecm r,r' <R \, .
\label{13}
\end{equation}
Substitution of Eq.~(\ref{15}) into Eq.~(\ref{4.1}) yields~\cite{gc91}
\begin{equation} 
     g(r,r';t) =\sum_{n=-\infty}^{\infty}u_n(r)u_n(r')M(k_n,t) \, ,
       \quad  r,r'<R \, ,
\label{16}
\end{equation}
where the $M$ function is defined as 
\begin{eqnarray}
    M(k_n,t)&=&
     \frac{\I}{2\pi}\int_{-\infty}^{\infty}\D k \, 
               \frac {\E ^{-\I k^2t}}{k-k_n}
          \nonumber\\
        &=& \frac{1}{2}w(\I y_n).
\label{16a}
\end{eqnarray}
The function $w$ is the complex error function~\cite{abramowitz}:
\begin{equation} 
      w(z)=\exp (-z^{2}){\rm erfc}(-\I z) \, . 
\end{equation}
The argument $y_n$ is given by 
\begin{equation}
      y_n = - \E ^{-\I \pi /4}k_nt^{1/2} \, .
         \label{17}
\end{equation}
One may write the above solutions in a form that exhibits explicitly its 
exponential and non-exponential contributions by writing them in terms of 
an expansion involving the complex poles of the fourth quadrant 
of the $k$-plane:
\begin{equation}
       g(r,r';t)=\sum_{p=1}^{\infty} \{ u_p(r) u_p(r') \E ^{-\I k_p^2t} 
    -[ u_p(r) u_p(r')M(-k_p,t) - u_p^*(r) u_p^*(r')M(-k_p^*,t)  ]  \} \, ,
         \label{24}
\end{equation}
where we have used that the poles of the third quadrant satisfy, from 
time reversal considerations, that $k_{-p}=-k_p^*$ and that 
$u_{-p}=u_p^*$. Also, we have made use of the following symmetry of the 
$w$ functions~\cite{abramowitz}:
\begin{equation}
        w(\I y_p)= 2 \E ^{y_p^2} -w(-\I y_p) \, ,
       \label{22a}
\end{equation}
which holds when the argument $y_p$ lies within the limits 
\begin{equation}
      \pi/2 < {\rm arg}\,y_p < 3\pi/2 \, .
\label{22b}
\end{equation}
This last condition is fulfilled by the proper poles of the fourth quadrant.

We note that, contrary to all previous expansions, the expansion~(\ref{16}) 
first yields an estimation of the background integral when $r,r'<R$, second
it is valid when $r,r'<R$, and third it includes the contribution of the 
anti-resonance states explicitly.

\section{Resonance expansions and the steepest descent method}
\label{sec:stdesc}

Resonance expansions of the wave function in coordinate or momentum 
representations, and of survival amplitudes arise naturally when calculating 
the corresponding integrals with the steepest descent technique. 
The important point is that, even though the poles are the same 
as in other approaches (say poles of the transmission 
amplitude in one-dimensional scattering),
their contributions may differ. In this paper, we shall illustrate the basic 
idea in coordinate representation
and for one-dimensional scattering off a potential with support $[0,d]$, 
although similar manipulations can be performed in other cases. Instead
of the wave number $k$, in this section we shall use $p/\hbar$, $p$ being 
the momentum.

Assume that the wave packet is
initially confined to the left of the potential so that it can be written, in 
terms of its initial momentum representation $\phi (p)$ and   
the scattering eigenstates of $H$, $\psi _p$, as 
\begin{equation}
\langle  x|\psi(t)\rangle =\int_{C} \D p \, 
        \langle x|\psi _p\rangle \E^{-\I Et/\hbar}\phi (p)  
\end{equation}
where 
\begin{equation}
     \langle x|\psi_p\rangle =h^{-1/2}\left\{
                    \begin{array}{ll}
                    \E ^{\I px/\hbar}+ R(p)\E ^{-\I px/\hbar}  &   x\le 0  \\
                    T(p)\E ^{\I px/\hbar}   &   x\ge d \, ,
\end{array}\right.
\end{equation}
and $C$ goes from $-\infty$ to $\infty$ above all singularities. The 
coefficients $R(p)$ and $T(p)$ are the reflection and transmission amplitudes 
for $p>0$, continued analytically elsewhere. The transmitted wave
packet may then be written as 
\begin{equation}
         \langle  x|\psi(t)\rangle = 
         h^{-1/2}\int_{C} \D p\, \E ^{\I xp/\hbar} \E ^{-\I Et/\hbar} 
                  T(p) \phi (p)  \, ,  
\end{equation} 
and the contour $C$ may be deformed along the steepest descent path. 

These integrals may be written in the form 
\begin{equation}
       {\cal I}=\int_{C} \D k\, \E ^{-\I (ak^2+kb)}g(k). 
\end{equation}
In simple cases, $g(k)$ is a meromorphic function. The saddle point of the 
exponent is at $k=-b/2a$, and the steepest descent path is the straight line
${\rm Im}(k)=-({\rm Re}(k)+b/2a)$. By completing the square, introducing 
the new variable $u$, 
\begin{equation}
         u=(k+b/2a)/f,\;\;\;\; f=(1-\I )(m\hbar/t)^{1/2}, 
\end{equation}
which is real on the steepest descent path and zero at the 
saddle point, and mapping the contour to the $u$-plane, 
the integral takes the form
\begin{equation}
    {\cal I}=\E ^{\I mx^2/\hbar t}f\int_{C_u} \D u\, \E ^{-u^2} G(u),
\end{equation}
where $G(u)\equiv g[k(u)]$. It is now useful to separate the pole 
singularities explicitly and write $G$ as 
\begin{equation}
   G(u)=\sum_j\frac{A_j/f}{u-u_j}+H(u), 
\end{equation}
where $A_j/f$ is the residue of $G(u)$ at $u=u_j$ and the remainder, $H(u)$,
is obtained by substraction. Note that $H(u)$ is an 
entire function if $G$ is meromorphic. 

The integral ${\cal I}$ is thus separated into two integrals, 
${\cal I}={\cal I}'+{\cal I}''$. The first one may be reduced to known 
functions by deforming the contour along the steepest descent 
path (real-$u$ axis) and taking proper care of the pole contribution,      
\begin{eqnarray}
      {\cal I}'&\equiv& \E ^{\I mx^2/\hbar t} \sum_j 
            A_j\int_{C_u} \D u\,\frac{\E ^{-u^2}}{u-u_0}
\nonumber\\
  &=&
        -\I \pi \E ^{\I mx^2/\hbar t} \sum_j A_j w(-u_j) \, .
\end{eqnarray}
The second integral, which involves the remainder $H$, 
must in general be evaluated numerically, 
\begin{equation}
         {\cal I}''\equiv \E ^{\I mx^2/\hbar t}f \int_{C_u} \D u\,H(u).
\end{equation} 
However, the computational effort is greatly reduced by deforming
the contour along the steepest descent path, too. When it is an entire 
function, it may be expressed as a series by expanding $H(u)$ around the 
origin and integrating term by term,        
\begin{eqnarray}
      {\cal I}''&=&\E ^{\I mx^2/\hbar t }f\pi^{1/2}\Bigg[ H(u=0)  \\
      &+&
            \left.\sum_{n=1}^\infty \frac{1\times3\times...\times(2n-1)}
              {2^n(2n)\!}H^{(2n)}(u=0)\right],
\nonumber
\end{eqnarray}
In practical applications, the first term alone gives already a very good
approximation. Note the basic role of the $w$-functions,
which may be thought of as the elementary transient mode
propagators of the Schr\"odinger equation. Long and short time formulae 
are easily obtained from asymptotic expansions of $w(z)$. 

For applications of this technique to describe different transient  
phenomena, see for example~\cite{BM96,DCM02,DMAG05}.

\section{Complex, effective Hamiltonians}

As we shall see in this section, truncation of resonance expansions enable 
us to understand the origin of complex, effective Hamiltonians.

If we apply the Hamiltonian to the Berggren expansion~(\ref{resonidBerggren}),
we obtain
\begin{equation}
     H=\sum_{n=1}^{N_{\rm b}} K_n^2 \, |K_n\rangle \langle K_n| +
        \sum_{n=1}^{N_{\rm B}} k_n^2 \, |k_n^+\rangle \langle ^+k_n| 
       + \int _{\Gamma _{\rm B}}\D k \, k^2 \, |k^+\rangle \langle ^+k| \, ,  
         \label{HresonidBerggren}
\end{equation}
Thus, the Hamiltonian can be split into a sum of bound, 
resonance and background contributions:
\begin{equation}
     H = H_{\rm bound} + H_{\rm resonance} + H_{\rm background} \, .
         \label{Hboupsec}
\end{equation}
In matrix notation, one can write this equation as a diagonal matrix,
\begin{equation}
     H= \left(  \begin{array}{ccclcccc}
                 K_1^2 & \,  & \, & \, &  \\
               \, & \ddots & \, & \, &    \\
               \, & \,  &  K_{N_{\rm b}}^2& \, &   \\
               \, & \,  & \,  & k_1^2 & \,  \\
               \, & \,  & \,  & \, & k_2^2 & \,  \\
               \, & \,  & \,  & \, & \, & \ddots &  \\
               \, & \,  & \,  & \, & \, & \, & k_{N_{\rm B}}^2 &  \\
               \, & \,  & \,  & \, & \, & \, &  \,          &  (k^2)_{\rm bg} 
                 \end{array}
         \right) . 
         \label{Hboupsec-matrix}
\end{equation}
When, for example, the first and second resonances are the most important
to the problem under consideration, we can neglect everything but the
contribution of those two resonances. The corresponding effective, complex 
Hamiltonian arises in a natural way:
\begin{equation}
     H= \left(  \begin{array}{ccclcccc}
                 k_1^2 & 0  \\
                 0 & k_2^2  \\
                 \end{array}
         \right) .
         \label{Hboupsec-eff}
\end{equation}
When double poles come into play, the effective Hamiltonians have non-diagonal
terms~\cite{MONDRAGON}. Note that, contrary to PT-symmetric Hamiltonians, the 
eigenvalues of this effective Hamiltonian are complex.



\section*{Acknowledgement}

RM acknowledges financial support from the Basque Government 
through reintegration fellowship No.~BCI03.96.

\vskip1cm

\begin{thebibliography}{99}

\bibitem{GAMOW} G.~Gamow: Z.~Phys.~{\bf 51} (1928) 204. 

\bibitem{CONDON} R.W.~Gurney and E.U.~Condon: Phys.~Rev.~{\bf 33} (1929) 127.

\bibitem{SIEGERT} A.F.J.~Siegert: Phys.~Rev.~{\bf 56} (1939) 750.

\bibitem{HUMBLET} J.~Humblet and L.~Rosenfeld: Nucl.~Phys.~{\bf 26} (1961) 529.

\bibitem{PEIERLS55} R.E.~Peierls, ``Interpretation and properties of 
propagators,'' in {\it The Proccedings of the 1954 Glasgow 
Conference on Nuclear and Meson Physics}, edited by E.H.~Bellamy and 
R.G.~Moorhouse (Pergamon Press, London and New York, 1955) p. 296-299.

\bibitem{PEIERLS59} R.E.~Peierls: Proc.~R.~Soc.~London, Ser.~A {\bf 253} 
(1959) 16.

\bibitem{ZELDOVICH} Ya.B.~Zeldovich: Sov.~Phys.~JETP~{\bf 12} (1961) 542.

\bibitem{BERGGREN} T.~Berggren: Nucl.~Phys.~A {\bf 109} (1968) 265.

\bibitem{gcp76} G.~Garc\'\i a-Calder\'on and R.~Peierls:
Nucl.~Phys.~A {\bf 265} (1976) 443.

\bibitem{ROMO} W.J.~Romo: J.~Math.~Phys.~{\bf 21} (1980) 311.

\bibitem{BG} A.~Bohm and M.~Gadella, {\it Dirac kets, Gamow 
Vectors, and Gelfand Triplets}, Springer Lectures Notes in 
Physics Vol.~348 (Springer, Berlin, 1989).

\bibitem{LIND} P.~Lind: Phys.~Rev.~C~{\bf 47} (1993) 1903.

\bibitem{VERTSE} T.~Vertse, R.J.~Liotta and E.~Maglione:
Nucl.~Phys.~A {\bf 584} (1995) 13.

\bibitem{BOLLINI} C.G.~Bollini, O.~Civitarese, A.L.~De Paoli
and M.C.~Rocca: J.~Math.~Phys.~{\bf 37} (1996) 4235.

\bibitem{TOLSTIKHIN} O.I.~Tolstikhin, V.N.~Ostrovsky and H.~Nakamura:
Phys.~Rev.~A~{\bf 58} (1998) 2077.

\bibitem{FERREIRA} L.S.~Ferreira and E.~Maglione: Chaos, Solitons \& 
Fractals~{\bf 12} (2001) 2697.

\bibitem{MONDRAGON} E.~Hernandez, A.~Jauregui and A.~Mondragon: 
Phys.~Rev.~A~{\bf 67} (2003) 022721.

\bibitem{CIVITARESE} O.~Civitarese: M.~Gadella, Phys.~Rep.~{\bf 396} (2004)
41.

\bibitem{MICHEL} N.~Michel, W.~Nazarewicz, J.~Okolowicz and M.~Ploszajczak:
Nucl.~Phys.~A~{\bf 752} (2005) 335c.

\bibitem{SANTRA} R.~Santra, J.M.~Shainline and C.H.~Greene, 
Phys.~Rev.~A~{\bf 71} (2005) 032703.

\bibitem{AJP02} R.~de la Madrid and M.~Gadella: Am.~J.~Phys.~{\bf 70} 
(2002) 626; {\sf quant-ph/0201091}.

\bibitem{BOHM} A.~Bohm, M.~Loewe and B.~van de Ven: Fortsch.~Phys.~{\bf 51} 
(2003) 551; {\sf quant-ph/0212130}.  

\bibitem{newton} R.G.~Newton: {\it Scattering Theory of Waves 
and Particles}, (Second edition, Springer-Verlag, New York, 1982) Chapter~12.

\bibitem{more} R.M.~More: Phys.~Rev.~{\bf A4} (1973) 1782.

\bibitem{gcb79} G.~Garc\'{\i}a-Calder\'on and M.~Berrondo: 
Lett.~Nuovo Cimento {\bf 26} (1979) 562.

\bibitem{gareev} F.A.~Gareev, M.H.~Gitzzatkulov and S.A.~Goncharov:
Nucl.~Phys.~{\bf 309} (1978) 381.

\bibitem{gc82} G.~Garc\'{\i}a-Calder\'on: Lett.~Nuovo Cimento {\bf 33} 
(1982) 253.

\bibitem{gcr86} G.~Garc\'{\i}a-Calder\'on and A.~Rubio: 
Nucl.~Phys.~A~{\bf 458} (1976) 560.

\bibitem{gc91} G.~Garc\'{\i}a-Calder\'on in {\it Symmetries in 
Physics}, edited by A.~Frank and K.B.~Wolf (Springer-Verlag, Berlin, 1992)
p.~252.

\bibitem{abramowitz} M.~Abramowitz and I.A.~Stegun: {\it Handbook of 
Mathematical Functions} (Dover Publications Inc., New York, 1972) p.~297-298.

\bibitem{BM96} S.~Brouard and J.G.~Muga: Phys.~Rev.~A {\bf 54}
(1996) 3055. 

\bibitem{DCM02} F.~Delgado, H.~Cruz and J.G.~Muga: J.~Phys.~A {\bf 35} 
(2002) 10377. 

\bibitem{DMAG05} F.~Delgado, J.G.~Muga, G.~Austing and 
G.~Garc\'\i a-Calder\'on: J.~Appl.~Phys.~{\bf 97} (2005) 013705.   







\end{thebibliography}
\end{document}